\documentclass[useAMS,usenatbib,usegraphicx,letterpaper]{mn2e}
\usepackage{graphicx}
\usepackage{amsmath}
\usepackage{epsfig}
\usepackage{subfigure} 

\newcommand{\be}{\begin{equation}}
\newcommand{\ee}{\end{equation}}

\newcommand{\pder}[2]{\frac{\partial #1}{\partial #2}}
\newcommand{\pdert}[1]{\pder{#1}{t}}
\newcommand{\bl}[1]{\mbox{\boldmath$ #1 $}}

\newcommand{\sph}{{\small SPH}}
\newcommand{\amr}{{\small AMR}}
\newcommand{\flash}{{\small FLASH}}
\newcommand{\ktscheme}{KT scheme}
\DeclareMathOperator{\sgn}{sgn}

\begin{document}

\title[]{Collisionless Stellar Hydrodynamics as an Efficient Alternative to N-body Methods}
\author[Nigel L. Mitchell, Eduard I. Vorobyov and Gerhard Hensler]{Nigel L. Mitchell$^{1}$\thanks{E-mail:nigel.mitchell@univie.ac.at}, Eduard I. Vorobyov$^{1,2}$, Gerhard Hensler$^{1}$ \\
$^{1}$University of Vienna, Dept. of Astrophysics, 1180 Vienna, Austria \\
$^{2}$Institute of Physics, Southern Federal University, Rostov-on-Don, 344090, Russia 
}

\maketitle

\begin{abstract}{}
The dominant constituents of the Universe's matter are believed to be collisionless in nature and thus their modelling in any self-consistent simulation is extremely important. For simulations that deal only with dark matter or stellar systems, the conventional N-body technique is fast, memory efficient, and relatively simple to implement. 
However when extending simulations to include the effects of gas physics, mesh codes are at a distinct disadvantage compared to \sph\ codes. Whereas implementing the N-body approach into \sph\ codes is fairly trivial, the particle-mesh technique used in mesh codes to couple collisionless stars and dark matter to the gas on the mesh, has a series of significant scientific and technical limitations. These include spurious entropy generation resulting from discreteness effects, poor load balancing and increased communication overhead which spoil the excellent scaling in massively parallel grid codes.

In this paper we propose the use of the collisionless Boltzmann moment equations as a means to model the collisionless material as a fluid on the mesh, implementing it into the massively parallel \flash\ \amr\ code. This approach which we term ``collisionless stellar hydrodynamics'' enables us to do away with the particle-mesh approach and since the parallelisation scheme is identical to that used for the hydrodynamics, it preserves the excellent scaling of the \flash\ code already demonstrated on peta-flop machines. 

We find that the classic hydrodynamic equations and the Boltzmann moment equations can be reconciled under specific conditions, allowing us to generate analytic solutions for collisionless systems using conventional test problems. We confirm the validity of our approach using a suite of demanding test problems, including the use of a modified Sod shock test. By deriving the relevant eigenvalues and eigenvectors of the Boltzmann moment equations, we are able to use high order accurate characteristic tracing methods with Riemann solvers to generate numerical solutions which show excellent agreement with our analytic solutions. We conclude by demonstrating the ability of our code to model complex phenomena by simulating the evolution of a two armed spiral galaxy whose properties agree with those predicted by the swing amplification theory.

\end{abstract}

\section{Introduction}
Galaxies are complex systems consisting of a multitude of components that often require different approaches for both analytical and numerical study. The gaseous component can be described with the classic hydrodynamics equations (CHE) thanks to frequent collisions between constituent particles that act to isotropise the local pressure. On the other hand, stellar and dark matter particles lack collisions which effectively means the local pressure is anisotropic and should be described by a stress tensor. Although this modification is rather straightforward to implement in the CHE \citep[see e.g.][]{Samland97,VT2006}, this approach has mostly been applied to analytical and semi-analytical studies of growing instabilities in stellar systems \citep[e.g.][]{BT87}.


When it comes to numerical simulations of collisionless galactic components, N-body methods have often been the method of choice \citep[e.g.][]{Athanassoula84}. One owes this to the relative ease with which they can be implemented and one can start modelling three-dimensional collisionless systems - often with rather small numbers of particles. In particular, large dark-matter-only N-body simulations \citep[e.g. the {\it Millennium simulation}, ][]{Springel2005} have led to significant advances in our understanding of cosmological structure formation, allowing some of the founding tenets of the current paradigm to be tested. 

As simulations have grown in complexity, incorporating the effects of baryonic matter, it has become necessary for theorists to develop a wider range of techniques to model the gaseous component. For this, there exist two dominant approaches; that of Smooth Particle Hydrodynamics (\sph) and alternatively, grid codes. Whilst both approaches agree for simple tests where analytic solutions are available, discrepancies are still often found when modelling more complex phenomena. \sph\ is very robust and memory efficient, allowing large cosmological runs to self-consistently model the effects of gas physics and collisionless matter \citep[see for example the {\it OWLS} and {\it GIMIC} simulations;][]{Schaye2009, Crain2009}. However, due to the higher shock capturing powers and more flexible refinement criteria of mesh codes, along with the tendency of \sph\ to suppress the formation of hydrodynamic instabilities and turbulent mixing \citep{Agertz2007,Mitchell2009}, grid codes present a powerful alternative. In particular, grid based schemes are better suited to modelling small-scale physical processes such as heat conduction and radiative transfer.

Unfortunately, when coupling the collisionless matter to the collisionally dominated baryonic matter, it begins to become clear that the N-body approach is not always ideally suited to grid codes. 
In \sph\ the gas is discretised into particles of a given mass, allowing the effects of gravity to be calculated using an N-body method in a very convenient self-consistent manner for all components. Grid codes however require the mapping of the collisionless particle masses to the mesh, from which the global density field can be used to obtain the gravitational potential. This can be done using the multigrid technique \citep{Fryxell2000} with either a nested grid or Adaptive Mesh Refinement (\amr). Alternatively a Fourier Transform or a hybrid of the two can accelerate this process. Once the gravitational acceleration is computed on the grid, it is then interpolated back to the particles. These are updated using a leapfrog method. This particle-mesh approach, a mapping of discrete particle properties to and from the mesh, represents a series of scientific and technical limitations to grid codes which degrade the physical reliability of their results and hampers the scalability of grid based simulations. These include:

\begin{itemize}

\item {\it Spurious generation of entropy} - The particle-mesh technique allows numerical noise to be introduced if insufficient particles are present in a given resolution element or ``cell.'' It has been argued by \citet{Springel2010} that in regions where the mesh is over-sampled relative to the number density of particles, then some cells may receive mass mapped from particles whilst their neighbours may not. This can lead to strong variations in the local density field when realistically the field should be smooth and continuous. This can drive spurious weak shocks and turbulence on small scales, generating an unphysical entropy excess. \citet{Springel2010} cites this as the origin of the higher entropy cores seen in galaxy cluster simulations when run using grid codes, compared to those run with \sph\ codes which have cuspy low entropy cores \citep[][]{Frenk1999,Mitchell2009}.

As actually these particles should represent a continuous mass field, this effect can be limited by mapping the discrete mass of a given particle over a larger region using some pre-defined kernel function \citep[for example the ``Cloud-in-cell'' technique;][]{Harlow1964}, this however leads to more technical problems. 


\item {\it Increased communication overhead} - In massively parallel programs, a large amount of effort is often put into minimising communication between different nodes over the network. Such communication is relatively time consuming, with latencies (the time taken to establish and terminate a communication) many orders of magnitude greater than the clock speed of the processor. As such, many calculations could be performed whilst waiting for a given group of processes to communicate, and if there is a list of different processes to communicate with, the time adds up.
Therefore any common communication scheme and grouping of messages, along with a reduction in the number of different nodes to which we must communicate with, presents a significant speed up in the code.

When particle masses are mapped using a kernel, there will inevitably be some overlap with neighbouring cells which may or may not reside on different nodes. Although communication over the network is natural for any parallel simulation, it is lower when dealing with a system which advects only fluxes through cell surfaces instead of a system which can map mass to any neighbouring cell, whether it is directly adjacent to a face or merely in contact with an edge or vertex of the cell. This is demonstrated in figure~\ref{fig:PMvsFlux} in which a schematic layout of two dimensional (2D) block of cells (as used in the \flash\ code) is shown along with the dashed outline of the neighbouring blocks which surround it. The internal cells in the block are shown in red and a surrounding layer four guardcells deep contains boundary data copied from neighbouring blocks. Grey cells are guardcells which do not need to be updated or communicated for the given scheme, whilst green cells are guardcells which need to be updated or exchanged with neighbours. A 2D hydrodynamic simulation (top panel) requires the communication of boundary data to compute the four fluxes through its cell faces with the grey corner guardcells not being used. The bottom panel however, shows how mapping of particle mass over a kernel of finite size can lead to particle properties being mapped to any of the surrounding guardcells. This data then needs to be communicated to the internal cells of the blocks to which they correspond. Thus in 2D the communication overhead is doubled, whilst in 3D it can increase by a factor of 3.3 (6 face fluxes / 20 edge and corner cells to which mass may be mapped). Sophisticated space filling curves can help optimise the distribution of blocks (or individual cells) across nodes so that as many adjacent neighbours as possible reside on the same processor, minimising inter-node communication. However increased overhead remains unavoidable, especially as we always remain bottle-necked by the slowest processor if the simulation is to proceed in lock-step.


\begin{figure}
\centering
\leavevmode 
\subfigure{ \includegraphics[trim=0cm 7cm 3cm 4cm, clip=true, height=8.4cm]{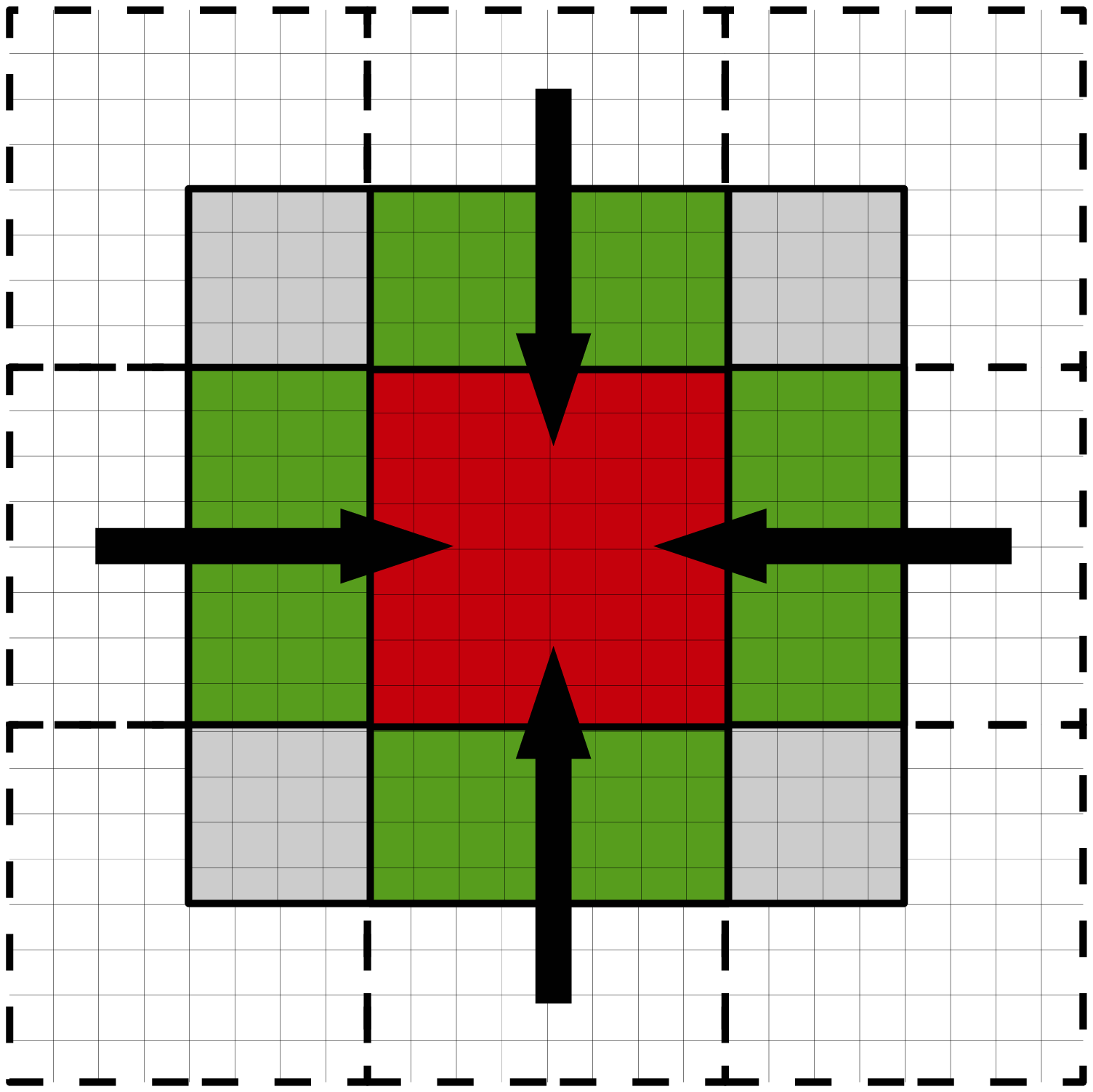} } \quad 
\subfigure{ \includegraphics[trim=0cm 7cm 3cm 4cm, clip=true, height=8.4cm]{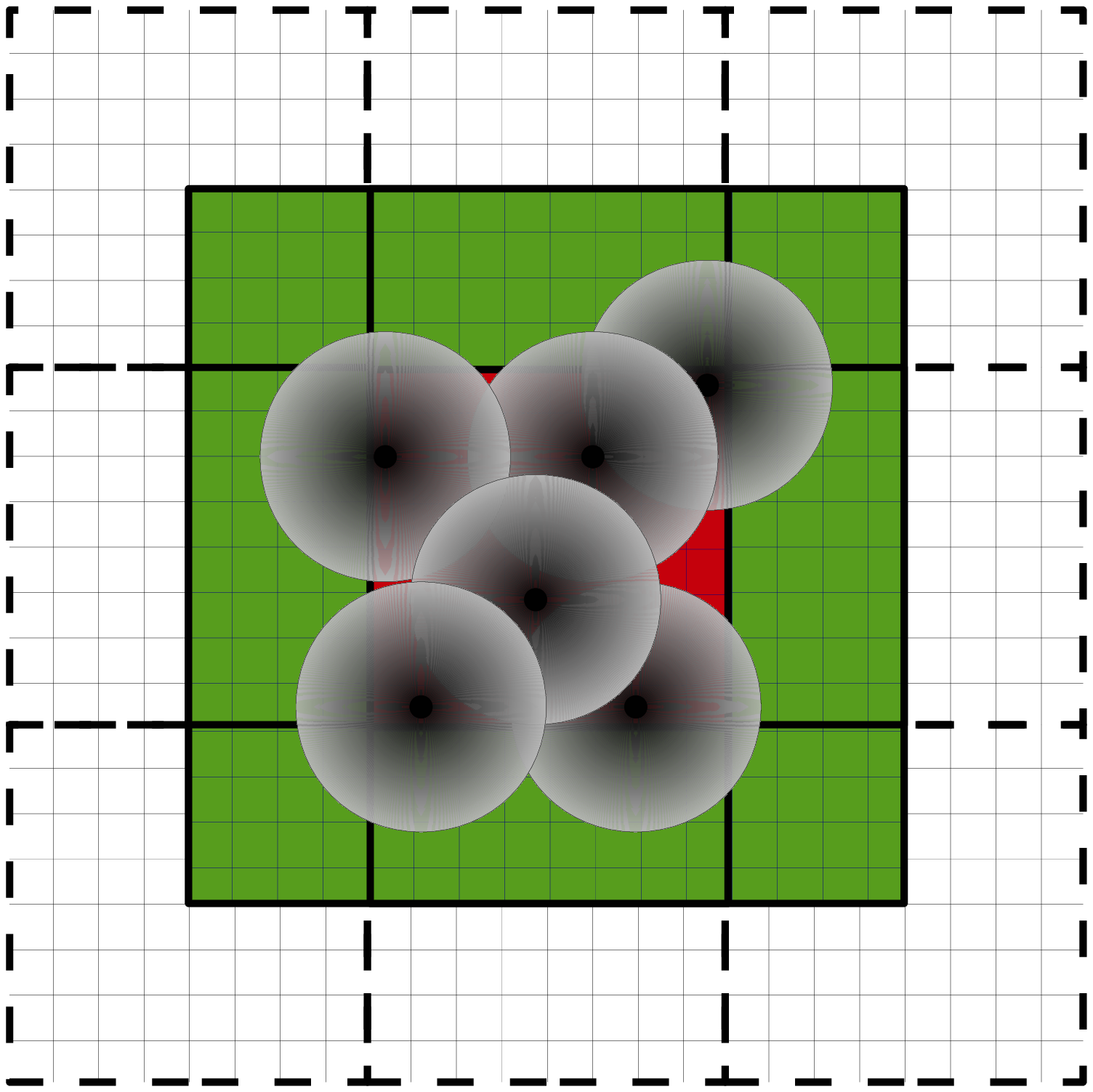} }
\caption{Diagrammatic representation of a block of cells and the required communication overhead for a flux based scheme (top panel) and a particle-mesh scheme (lower panel). The internal cells of the block are shown in red.  Surrounding guardcells which contain boundary data mapped from the internal cells of neighbouring blocks (visible as dashed squares) are shown in grey. Those guardcells which need to be updated after every step for the given scheme are highlighted in green. In the top panel, only flux data through the cell faces needs calculating and therefore only guardcell data directly adjacent to these surfaces needs to be communicated. Thus in 2D, only four of the eight neighbouring blocks need to be communicated with. However, the particle-mesh algorithm (bottom panel) which maps particle properties to the mesh, over a finite region with some Kernel, allows data to be mapped into all of the guardcells. Therefore it is necessary to perform a full guardcells exchange and to update the corresponding internal cells of all neighbouring blocks. This results in a much greater communication overhead.
 }
\label{fig:PMvsFlux}
\end{figure}

\item {\it Poor load balancing} - The use of particles in conjunction with a mesh also presents a load-balancing issue as many forms of astrophysical structure formation result in the natural concentration of the majority of mass in dense compact regions. This leads to too many particles accumulating in one or two cells whilst the remaining cells have few or none, even if the initial particle distribution is fairly uniform. Ultimately, the mass field ends up poorly sampled in the under-dense regions of the simulation whilst the mapping of properties to and from the mesh is performed by a limited number of cells within the densest regions. As already mentioned, the hydrodynamics requires that as many neighbouring cells as possible be stored on the same processor. Therefore these heavily occupied dense cells are all stored on the same few processors leading to extremely poor load balancing. Although distributing these dense blocks more evenly would improve the load balancing for the particle-mesh algorithm, it would ruin the load balancing for the hydrodynamics, creating an impossible system to optimise.

\end{itemize}

Fortunately the use of the collisionless Boltzmann moment equations allows us to overcome all of these limitations since it allows the stellar and dark matter components to be represented as a collisionless fluid in a near identical manner to the standard baryonic gas. This removes the need to develop a separate parallelisation scheme for the particle-mesh approach and allows us to use the same well parallelised hydrodynamics scheme. Instead of communicating guardcell data for the gas and stellar material separately, the data can now be grouped and communicated all at once, minimising the latency overhead and maximising bandwidth usage. Ideal load-balancing is also achieved as the time taken to calculate the inter-cell fluxes is independent of the amount of collisionless mass within a cell. Thus processors will not be left idle whilst others calculate, provided that the cells are evenly distributed across processors. 
Given that the hydrodynamics in the \flash\ \amr\ code we use has been shown to scale well for tens of thousands of processors \citep{Antypas2006}, then by using the same parallelisation scheme for the collisionless material, we naturally preserve this scaling. In an era of increasingly large and complex simulations, this scalability is vital in keeping performance in-line with scientific requirements.

Most importantly, as the density of the collisionless material is now represented in each cell continuously, this removes the potential for spurious generation of entropy through discreteness effects introduced when mapping particle properties to the mesh. It also removes limitations involved when converting gas into stars of discrete masses - problematic if cells contain insufficient mass but yet have densities that satisfy star formation criteria.


We present in this paper our new implementation of the Boltzmann moment equations which allows collisionless material to be modelled as a collisionless fluid on the mesh. A major application of this technique will be in modelling the behaviour of large collections of stars, such as those in a galaxy, as a collisionless fluid in much the same way that we model gas hydrodynamics in grid codes. We will therefore refer to it from here on as ``collisionless stellar hydrodynamics.''
We begin in \S~\ref{sec:boltzmanneqns} with a derivation of the zeroth, first and second order moment equations of the collisionless Boltzmann equation and then proceed in \S~\ref{sec:numericalschemes} to outline the \flash\ hydrodynamic code (\S~\ref{sec:flashcode}) and the different numerical schemes we use to numerically integrate the equations. These include both a simpler Riemann solver free technique (\S~\ref{sec:ktscheme}) and a more sophisticated Riemann solver scheme with characteristic tracing (\S~\ref{sec:riemannsolver}). 

In \S~\ref{sec:tests} we outline our discovery of how standard hydrodynamic test problems for which analytic solutions are known, can be modified under special conditions to reproduce the results of a collisionless fluid. This enables us to use the Sod-shock-tube test problem to test our results, the findings of which are given in \S~\ref{sec:sodshock}. We continue our tests with both a pressure-free spherical collapse test (\S~\ref{sec:sphercollapse}) and an anisotropic collapse with non-isotropic dispersion tensor in \S~\ref{sec:anisocollapse}.
As a more complex test of our collisionless stellar hydrodynamics, in \S~\ref{sec:diskgal} we track the growth of a two armed spiral in a stellar disc, contrasting the resulting structure
with the predictions of the swing amplification theory.

We then conclude our findings with an overview of our results, a discussion of the runtime performance of our code and the future implications of our work in \S~\ref{sec:summary}

\section{Boltzmann moment equations}
\label{sec:boltzmanneqns}
The dynamics of collisionless particles are fundamentally described by the
collisionless Boltzmann equation for the distribution function of  particles
in the six dimensional position-velocity phase space $f(t,{\bl x},{\bl v})$
\begin{equation}
   \label{Boltzmann} 
   \pdert{f} + v_i \cdot \pder{f}{x_i} + 
      a_i \cdot \pder{f}{v_i} = 0 \; ,
\end{equation}
where $a_i\equiv dv_i/dt$ is the acceleration due to gravity and curvilinear
terms (in non-Cartesian coordinate systems). In this paper, we use Cartesian coordinates and
hence $i\equiv(x,y,z)$.
Equation~(\ref{Boltzmann}) is a six-dimensional, time-dependent partial differential 
equation and although its direct numerical solution is possible \citep[see, e.g.][]{Yoshikawa2012}
high resolution simulations are prohibitively expensive even for the most powerful supercomputers. 
N-body methods circumvent this problem by solving a system of ordinary differential equations
for the dynamics of individual particles, which on galactic scales often represent a cluster 
of objects rather than individual physical entities. In contrast to this {\it discrete}
mechanics approach,
a {\it continuous} mechanics approach is based on taking the first three velocity moments of 
equation~(\ref{Boltzmann}). More specifically, equation~(\ref{Boltzmann})
is multiplied by ten quantities $m, m v_i, v_i v_j$, where $i\equiv(x,y,x)$, 
and integrated over the velocity space $d^3{\bl v}\equiv dv_{\rm x} dv_{\rm y} dv_{\rm z}$
to obtain partial differential equations for the volume density
$\rho =\int m f d^3{\bl v}$, the mean velocity $\bl{u}=\rho^{-1} \int m f \,{\bl v}\, d^3{\bl v}$, 
and velocity dispersions $\sigma^2_{ij}=\rho^{-1}\int m f\, (v_i-u_i)(v_j-u_j)\,d^3{\bl v}$
of collisionless particles. 

The resulting continuity, momentum, and velocity dispersion equations, named Boltzmann moment equations (BME), are as follows, \\
\begin{eqnarray}
\label{contin}
\pder{\rho}{t} & + & \pder{}{x_i}\left( \rho u_i \right) = 0 \; , \\
\label{moment}
\pder{}{t} \left( \rho u_i \right) &+& \pder{}{x_j} \left( \rho u_i \cdot u_j + \Pi_{ij} \right)
-\rho a_i =0 \; , \\
\label{dispers}
\pder{E_{ij}}{t} &+& \pder{}{x_k}\left( E_{ij} u_k + \rho \sigma_{jk}^2 u_i + \rho 
\sigma_{ik}^2 u_j  \right) \nonumber \\
 &-& \rho u_i a_j - \rho u_j a_i = 0 \; .
\end{eqnarray}
Here, $E_{ij}=\Pi_{ij} + \rho u_i u_j$ and $\Pi_{ij}=\rho \sigma_{ij}^2$ is the 
symmetric velocity dispersion tensor. All indices obey the usual Einstein summation rule.
Equations~(\ref{contin})-(\ref{dispers}) are closed by using the usual 
zero-heat-flux approximation and setting the third-order moments to zero, 
$Q_{ijk}=\rho^{-1}\int f\,(v_i-u_i)(v_j-u_j)(v_k-u_k)\,d^3{\bl v}=0$.
We discuss the applicability of this approximation in Appendix B.

Equations~(\ref{contin}) and (\ref{moment}) are in fact the Jeans equations \citep[e.g.][]{BT87},
from which the classic hydrodynamics equation are derived assuming a diagonal and 
isotropic form of the stress tensor $\Pi_{ij}$ and introducing the gas pressure 
$P=1/3 \sum \limits_i \rho 
\sigma_{ii}^2$ and internal energy per unit volume $\varepsilon=1/2\sum \limits_j \rho \sigma_{jj}^2$,
where $i$ and $j$ are the number of translational and total (translational plus internal) 
degrees of freedom.
In the Boltzmann moment equations (BME), no simplifying assumptions regarding the form of the 
stress tensor are made.
Due to this reason, every component of $\Pi_{ij}$ needs to be evolved in time separately.
Equation~(\ref{dispers}) does that for the quantity $E_{ij}=\Pi_{ij}+\rho u_i u_j$, 
which can be regarded as an analogue to the total energy $\varepsilon+1/2 \rho |{\bl u}|^2$ 
in the CHE. In total, one has 10 equations for ten unknown quantities: $\rho$, 
$\rho u_i$, and $\rho \sigma_{ij}^2$. 
We note that equations~(\ref{contin})-(\ref{dispers}) can be convolved to the usual CHE if 
an isotropic and diagonal form of $\Pi_{ij}$ is assumed (see Appendix A).

The fundamental similarity of the BME with the CHE makes it straightforward to couple
collisionless (stars, dark matter) and collisional (gas, dust) galactic subsystems
in one grid-based code. Below, we provide a detailed implementation of two numerical solvers which can accurately follow the behaviour of a collisionless fluid in the \flash\ \amr\ code.

\section{Numerical schemes}
\label{sec:numericalschemes}

A full solution of the collisionless Boltzmann equation requires a detailed knowledge of the velocity phase space distribution function for a collisionless fluid. Unfortunately no known solution is currently available and therefore the first scheme we choose to implement is one which neither requires any detailed knowledge of the systems equation of state, nor any consideration of the behaviour of the different entropy and pressure waves around cell interfaces as would be used by more sophisticated Riemann solver schemes. 
For this purpose we adopt the unsplit solver of \citet{KurganovTadmor2000} for hyperbolic partial differential equations (\ktscheme\ hereafter). The scheme is second order accurate in both space and time using a simplistic Runge-Kutta midpoint technique to ensure that the solution is second order accurate in time. This removes the need to deconvolve the moment equations into their separate eigenvectors within the Riemann interaction fan. Despite being Riemann solver free, leading to an averaging over interactions within the Riemann fan, when modelling a conventional ideal gas \citet{KurganovTadmor2000} have shown that it possess very low numerical diffusion and can accurately capture sharp shocks and contact discontinuities. This makes it an ideal means to test the reliability of higher order more sophisticated schemes and as it is also highly extensible, higher order moments of the Boltzmann equation can be readily incorporated. 

The second scheme we implement involves a more accurate modified state reconstruction using a new set of eigenvectors so that Riemann solvers can be directly applied. Unlike the \ktscheme, the Riemann solver approach achieves second order accuracy in time by reconstructing the time averaged fluid properties (primitive or characteristic) using more detailed knowledge of the eigenvector structure of the fluid equations. This grants the shock superior resolving powers, allowing sharper changes in the fluid variables to be tracked with less numerical diffusion. The scheme is also found to be relatively efficient compared to more dissipative schemes. As far as the authors are aware, this represents the first time that direct solution of the Riemann problem for an anisotropic collisionless fluid has been achieved, with previous efforts focussing on either the use of less complex schemes such as the aforementioned \ktscheme\ or alternatively more complete direct integration of the collisionless Boltzmann equation which represents a more significant computational overhead.
%


\subsection{The FLASH Code}
\label{sec:flashcode}

\flash\ was developed at the Alliance Centre for Astrophysical Thermonuclear Flashes as part of a DOE grant. Originally intended for the study of X-ray bursts and supernovae, it has since been adapted for many astrophysical conditions and now includes modules for relativistic hydrodynamics, thermal conduction, radiative cooling, magneto-hydrodynamics, thermonuclear burning, self-gravity and particle dynamics via a particle-mesh approach. In this study we use FLASH version 3.2.

The code is massively parallel, scaling well over more than tens of thousands of processors \citep[][]{Antypas2006}, and is highly modular enabling several different solvers for the hydrodynamics and stellar dynamics to be introduced. 

The code features an Oct-tree structure with the simulation being divided into ``blocks'', each of which contains a fixed number of cells; $N_{cells}=16$ in each dimension for our simulations. In addition to these cells, an additional four ``guardcells'' either side of the internal cells are required, which contain a copy of the properties of the neighbouring blocks in-case they are stored on a different node.

The code is capable of adaptive mesh refinement, increasing the resolution in regions where a user defined refinement criteria is satisfied and conversely removing refinement in regions where excessive refinement is both unnecessary and wasteful. Refinement is achieved by sub-dividing each ``parent'' block into $2^{N_{DIM}}$ ``child'' blocks. Those at the highest level of refinement for their region of space are termed ``leaf blocks.'' The code enforces a maximum jump of two between levels of refinement. At refinement level $l$, a fully refined AMR grid will have a total of $N_{cells}^{(2l - 1)}$ cells on a side.

To identify regions of rapid flow change, \flash's refinement and de-refinement criteria can incorporate the adapted \citet{Loehner1987} error estimator. This calculates the modified second derivative of the desired variable, normalised by the average of its gradient over one cell. With
this applied to the density as is common place, we impose maximum and minimum levels of refinement; $l_{max}$ and $l_{min}$. 

\subsection{The Kurganov and Tadmor Scheme}
\label{sec:ktscheme}

We implement into the \flash\ \amr\ code the semi-implicit numerical advection scheme of \citet{KurganovTadmor2000}. We briefly outline the scheme below but refer readers to their paper for a more in-depth discussion. In conjunction with the semi-implicit scheme, we use a Runge-Kutta integration scheme for the temporal integration using the following time derivative for the fluid variables, $u$ \citep[eqn 4.2 in][]{KurganovTadmor2000},
\begin{equation}
\frac{d}{dt} u_i(t) = - \frac{H_{i+1/2}(t) - H_{i-1/2}(t) }{\Delta x} \, ,
\end{equation}
\noindent
where the numerical fluxes are defined as,
\begin{equation}
\renewcommand{\arraystretch}{2.}
\begin{array}{lcl}
H_{i+1/2} & = & \frac{ f\bigl(u^{+}_{i+1/2}(t)\bigr)+f\bigl(u^{-}_{i+1/2}(t)\bigr)}{2} \\ \; & \; & -  \frac{a_{i+1/2}(t)}{2} \left( u^{+}_{i+1/2} - u^{-}_{i+1/2} \right) \; .
\end{array}
\renewcommand{\arraystretch}{1.}
\end{equation}

Here $u_{i-1/2}$ and $u_{i+1/2}$ represent the fluid variable values at the left and right side of cell $i$ with the indices $+$ and $-$ representing the interpolated values to the left and right sides of the respective cell interfaces. The fluxes $f(u)$ are those taken from the derived BME (\ref{contin}, \ref{moment} and \ref{dispers}). The quantity $a$ represents the ``spectral frequency'' of the system;

\begin{equation}
a_{i+1/2} = max \left\{ \rho\left(\frac{\partial f}{\partial u}(u^{-}_{i+1/2}) \right) , 
\rho\left(\frac{\partial f}{\partial u}(u^{+}_{i+1/2}) \right)   \right\} \; .
\label{eqn:spectralfreq}
\end{equation}

As this is the speed with which information is propagated, it is unsurprising that this equates to the effective sound speed of the collisionless fluid. This can be computed component wise with no loss of numerical accuracy, or alternatively as we show in \S~\ref{sec:tests} that it is possible to reconcile the BME with the CHE, we can use the conventional sound speed formula (\ref{eqn:soundspeed}) in a directionally dependant manner. This makes use of a value for $\gamma =3$ and the dispersion tensor $\Pi_{jk}$ which can be treated as a directionally dependant pressure $P_{jk}=\Pi_{jk}=\rho\sigma_{jk}^2$.

\begin{equation}
a_{jk} = \sqrt{\frac{\gamma \Pi_{jk}}{\rho}} = \sqrt{\gamma \sigma^{2}_{jk}} \; .
\label{eqn:soundspeed}
\end{equation}
\noindent

The scheme can easily be extended to arbitrary orders of temporal and spatial accuracy \citep[see ][]{KurganovLevy2000}, however we choose to implement a second order accurate Runge-Kutta solver, providing a good compromise between accuracy and computational cost. Higher order schemes would require additional guardcell updates at the many midpoints, increasing communication overhead, whereas the second order accurate scheme can perform the update using just the information in locally stored guardcells. 

The only modification which needs to be made in order for the scheme to function well with the adaptive mesh, is that flux conservation be checked at refinement boundaries. Fluxes calculated for higher resolution cells are considered to be more accurate than those at lower levels of refinement. Therefore the mean flux from the four higher resolution cells on a refinement boundary is calculated and passed to the neighbouring block at the coarser level of refinement to be used in place of the locally calculated flux.

\subsection{Accurate Solution of the Riemann Problem}
\label{sec:riemannsolver}

We implement our Riemann solver based collisionless stellar hydrodynamics module into \flash\ using the MUSCL-Hancock dimensionally unsplit scheme which has already been successfully implemented for the CHE including Magneto-hydrodynamics \citep{Dongwook2009}. We adopt an unsplit approach over a less memory intensive split version \citep[e.g. PPM of][]{Colella1984} given its superior ability to preserve flow symmetries. This is necessary given the highly non-linear nature of the BME which would otherwise quickly exaggerate any anisotropies. As an aside, we also note that we have significantly reduced the memory consumption of the default MUSCL-Hancock scheme provided with \flash\ through more efficient memory management, ensuring that the increased memory overhead of using the unsplit scheme is equivalent to only a single extra AMR block. The results for the reduced memory version are of course bit-wise identical to the default release.

Our collisionless scheme differs from the standard hydrodynamics scheme primarily through our determination of the new eigenvalues and eigenvectors which are required for the characteristic tracing step. This allows for the calculation of the time averaged fluid properties which can then be used to calculate the second order accurate fluxes using a conventional Riemann solver such as the approximate HLL family of schemes including the HLLC solver \citep[see, e.g.][]{Toro1999, Toro1994, Quirk1994} or more accurate Roe scheme \citep[][]{Roe1981}.

We begin by expressing the zeroth, first and second order moments of the Boltzmann equation (see equations \ref{contin}, \ref{moment} and \ref{dispers}) in compact matrix notation form (neglecting gravity for the moment),
\begin{equation}
{\bf U}_t + {\bf F}({\bf U})_x + {\bf G}({\bf U})_y + {\bf H}({\bf U})_z = 0 \; ,
\end{equation}

\noindent
with the conserved advection quantities ${\bf U}$ and their corresponding fluxes {\bf F}({\bf U}), {\bf G}({\bf U}) and {\bf H}({\bf U}) in the respective $x$, $y$ and $z$-directions being,

\begingroup
    \fontsize{8pt}{10pt}\selectfont
\begin{equation}
\centering
\begin{array}{cc}

{\bf U} = \left[ 
\begin{array}{c}
 \rho \\
 \rho v_x \\
 \rho v_y \\
 \rho v_z \\
 E_{xx} \\
 E_{yy} \\
 E_{zz} \\
\end{array} 
\right]  ,

& 

{\bf F}({\bf U}) 
= \left[ 
\begin{array}{l}
 \rho v_x \\
 \rho v_x v_x + \rho\sigma_{xx}^2\\
 \rho v_y v_x \\
 \rho v_z v_x \\
 E_{xx} v_x + 2\rho\sigma_{xx}^2 v_x\\
 E_{yy} v_x \\
 E_{zz} v_x \\
\end{array} 
\right]  ,

\\

{\bf G}({\bf U}) = \left[ 
\begin{array}{l}
 \rho v_y \\
 \rho v_x v_y \\
 \rho v_y v_y + \rho\sigma_{yy}^2\\
 \rho v_z v_y \\
 E_{xx} v_y \\
 E_{yy} v_y + 2\rho\sigma_{yy}^2 v_y\\
 E_{zz} v_y \\
\end{array} 
\right]  ,

&

{\bf H}({\bf U}) = \left[ 
\begin{array}{l}
 \rho v_z \\
 \rho v_x v_z \\
 \rho v_y v_z \\
 \rho v_z v_z + \rho\sigma_{zz}^2 \\
 E_{xx} v_z \\
 E_{yy} v_z \\
 E_{zz} v_z + 2\rho\sigma_{zz}^2 v_z\\
\end{array} 
\right]
,
\end{array}
\label{conservedfluxes}
\end{equation}
\noindent
\endgroup
where $E_{jk} = \rho v_{jk}^2 + \rho\sigma_{jk}^2$.
For simplicity, in this paper we assume a diagonalised velocity dispersion tensor in the above flux vectors, neglecting off diagonal terms, ie. $ \sigma_{jk} = 0 \; \forall \; j \neq k $. As the dimensions of the matrix and therefore complexity of the problem grow quickly with the increasing order of moments that we consider, we leave the tracking of the off diagonal terms and higher order moments to our next paper.

By taking the conservative equations and expanding the derivatives, it is possible to express equivalent advection equations for the primitive quantities; density, bulk velocity and velocity dispersion:

\begingroup
    \fontsize{8pt}{10pt}\selectfont
\begin{equation}
\centering
\begin{array}{cc}

{\bf W} = \left[ 
\begin{array}{c}
 \rho \\
 v_x \\
 v_y \\
 v_z \\
 \rho\sigma_{xx}^2 \\
 \rho\sigma_{yy}^2 \\
 \rho\sigma_{zz}^2 \\
\end{array} 
\right],

& 

\tilde{{\bf F}}({\bf W}) = \left[ 
\begin{array}{l}
 \rho v_x \\
 v_x v_x + \sigma_{xx}^2\\
 v_y v_x \\
 v_z v_x \\
 \rho\sigma_{xx}^2 v_x + \gamma \rho\sigma_{xx}^2 v_x\\
 \rho\sigma_{yy}^2 v_x \\
 \rho\sigma_{zz}^2 v_x \\
\end{array} 
\right],

\\

\tilde{{\bf G}}({\bf W}) = \left[ 
\begin{array}{l}
 \rho v_y \\
 v_x v_y \\
 v_y v_y + \sigma_{yy}^2\\
 v_z v_y \\
 \rho\sigma_{xx}^2 v_y \\
 \rho\sigma_{yy}^2 v_y + \gamma \rho\sigma_{yy}^2 v_y\\
 \rho\sigma_{zz}^2 v_y \\
\end{array} 
\right],

&

\tilde{{\bf H}}({\bf W}) = \left[ 
\begin{array}{l}
 \rho v_z \\
 v_x v_z \\
 v_y v_z \\
 v_z v_z + \sigma_{zz}^2 \\
 \rho\sigma_{xx}^2 v_z \\
 \rho\sigma_{yy}^2 v_z \\
 \rho\sigma_{zz}^2 v_z + \gamma \rho\sigma_{zz}^2 v_z\\
\end{array} 
\right],
\end{array}
\label{primitivefluxes}
\end{equation}
\noindent
\endgroup
where $\gamma=3$ (see \S~\ref{sec:tests}).

For brevity we now consider just the $x$-directional flux vectors in the conserved quantities, ${\bf F}({\bf U})$ and primitive variables, $\tilde{{\bf F}}({\bf W})$. 
As the components of the primitive flux vectors $\tilde{f}_i$ are each functions of the individual primitive quantities $u_i$, the Jacobian of the system in the x-direction, $\tilde{{\bf A}}({\bf W})_x$, can be expressed in terms of the partial derivatives of $\tilde{{\bf F}}({\bf W})$ with respect to ${\bf W}$:   

\begin{equation}
\tilde{{\bf A}}({\bf W}) = \partial \tilde{{\bf F}}/\partial {\bf W} = 
\left[
\begin{array}{ccc}
\partial \tilde{f}_1 / \partial w_1 & \ldots & \partial \tilde{f}_1 / \partial w_n \\
\vdots   & \ddots & \vdots \\
\partial \tilde{f}_n / \partial w_1 & \ldots & \partial \tilde{f}_n / \partial w_n \\
\end{array}
\right] \; .
\end{equation}

The eigenvalues $\lambda_i$ of the system are then obtained by solving for the roots of the polynomial,

\begin{equation}
| \tilde{{\bf A}} - {\bf \lambda I} | = 0 \; ,
\end{equation}

\noindent
where $I$ is the identity matrix. Physically, the eigenvalues represent the speeds of signal propagation within the system and are thus important for determining the extent to which fluid bounding a cell interface affects the flux through it over the time step. Unlike a collisionally dominated fluid which has five eigenvalues, the (diagonalised) collisionless fluid has seven to account for the anisotropy in the dispersion tensor (For a non-diagonalised version, this increases to ten). We determine the eigenvalues for the $x$-direction to be

\begin{equation}
 \lambda = [v_x-a_{xx}, v_x, v_x, v_x, v_x+a_{xx}, v_x, v_x]^T \; ,
\label{eigenvalues}
\end{equation}
\noindent
with $a_{xx}$ being the sound speed in the $x$-direction. The directionally dependant sound speed, $a_{jk}$, is in close analogy to that of a standard collisionally dominated gas,
\begin{equation}
a_{jk} = \sqrt{ \frac{\gamma P_{jk}}{\rho} } = \sqrt{ \gamma \sigma^2_{jk} } \; .
\end{equation}
\noindent

The right eigenvectors of the Jacobian, ${\tilde{\bf R}}^{(i)}$, are those that obey the relation $ {\tilde {\bf A}} {\tilde {\bf R}}^{(i)} = {\bf \lambda}_i {\tilde {\bf R}}^{(i)} $ whilst the left eigenvectors of the Jacobian, ${\tilde {\bf L}}^{(i)}$, are those which satisfy ${\tilde {\bf L}}^{(i)} {\tilde {\bf A}} = {\bf \lambda}_i {\tilde {\bf L}}^{(i)}$. Using the Jacobian for the primitive variable matrix equations,

\begin{equation}
\tilde{{\bf A}}({\bf W})_x = \left[
\begin{array}{ccccccc}
v_x & 0 & 0 & 0 & 0 & 0 & 0 \\
0 & v_x & 0 & 0 & \frac{1}{\rho} & 0 & 0 \\
0 & 0 & v_x & 0 & 0 & 0 & 0 \\
0 & 0 & 0 & v_x & 0 & 0 & 0 \\
0 & \rho a_{xx}^2 & 0 & 0 & v_x & 0 & 0 \\
0 & 0 & 0 & 0 & 0 & v_x & 0 \\
0 & 0 & 0 & 0 & 0 & 0 & v_x
\end{array}
\right] \; ,
\label{jacobian}
\end{equation}

\noindent
and the equivalent eigenvalues (\ref{eigenvalues}), the eigenvector matrices are found to be:

\begin{equation}
\tilde{{\bf R}}_x = \left[
\begin{array}{ccccccc}
\rho   & 0 & 0 & 1 & 0 & 0 & \rho \\
-a_{xx} & 0 & 0 & 0 & 0 & 0 & a_{xx} \\
0 & 0 & \frac{-1}{\rho} & 0 & 0 & 0 & 0 \\
0 & 0 & 0 & 0 & \frac{1}{\rho} & 0 & 0 \\
\rho a_{xx}^2 & 0 & 0 & 0 & 0 & 0 & \rho a_{xx}^2 \\
0 & \frac{-1}{\rho} & 0 & 0 & 0 & 0 & 0 \\
0 & 0 & 0 & 0 & 0 & \frac{1}{\rho} & 0 
\end{array}
\right] \; ,
\label{righteigvec}
\end{equation}

\begin{equation}
\tilde{{\bf L}}_x = \left[
\begin{array}{ccccccc}
0 & \frac{-1}{2a_{xx}} & 0 & 0 & \frac{1}{2\rho a_{xx}^2} & 0 & 0 \\
0 & 0 & 0 & 0 & 0 & -\rho & 0 \\
0 & 0 & -\rho & 0 & 0 & 0 & 0 \\
1 & 0 & 0 & 0 & \frac{ -1}{a_{xx}^2} & 0 & 0 \\
0 & 0 & 0 & \rho & 0 & 0 & 0 \\
0 & 0 & 0 & 0 & 0 & 0 & \rho \\
0 & \frac{1}{2a_{xx}} & 0 & 0 & \frac{1}{2\rho a_{xx}^2} & 0 & 0 
\end{array}
\right] \; .
\label{lefteigvec}
\end{equation}




With the general form of the eigenvalues and corresponding eigenvectors calculated, we can implement them into the MUSCL-Hancock scheme in-line with the approach taken by \citet[][]{Stone2008} which we outline below. First we compute the left, right and central differences of the primitive variables using the cell centred values within each cell, where $i$ is the cell index.
\begin{eqnarray}
\delta {\bf w}_{L,i} = {\bf w}_{i} - {\bf w}_{i-1} \;\; , \\
\delta {\bf w}_{C,i} = ({\bf w}_{i+1} - {\bf w}_{i-1})/2 \; , \\
\delta {\bf w}_{R,i} = {\bf w}_{i+1} - {\bf w}_{i} \; .
\end{eqnarray}
\noindent
We then map them onto the characteristic variables using the left eigenvectors calculated using (\ref{lefteigvec}), where ${\tilde {\bf L}}({\bf w})_i$ is the matrix corresponding to the $i^{th}$ row of the left eigenvector matrix, and the primitive variables ${\bf w}_{i}$ are defined at the cell centres,
\begin{eqnarray}
\delta {\bf u}_{L,i} = {\tilde {\bf L}}({\bf w}_{i}) \cdot \delta {\bf w}_{L,i} \; , \\
\delta {\bf u}_{C,i} = {\tilde {\bf L}}({\bf w}_{i}) \cdot \delta {\bf w}_{C,i} \; , \\ 
\delta {\bf u}_{R,i} = {\tilde {\bf L}}({\bf w}_{i}) \cdot \delta {\bf w}_{R,i} \; .
\end{eqnarray}
\noindent
As the scheme is higher than first order, the reconstruction of the time averaged variables at the cell interfaces, can lead to the formation of new non-physical local maxima or decreasing minima. To prevent their formation which otherwise leads to spurious post shock oscillations, we apply monotonicity constraints to the characteristic differences, ensuring that the time averaged variables are total variation diminishing \citep[TVD, see ][]{LeVeque2002},
\begin{equation}
\delta {\bf u}_i^m = \sgn (\delta {\bf u}_{C,i}) \min(2|\delta {\bf u}_{L,i}|,|\delta {\bf u}_{C,i}|,2|\delta {\bf u}_{R,i}|) \; .
\end{equation}
\noindent
The monotonised differences are then projected back onto the primitive variables, where ${\tilde {\bf R}}({\bf w}_i)$ is the matrix corresponding to the $i^{th}$ column of the right eigenvector matrix (\ref{righteigvec}), 

\begin{equation}
\delta {\bf w}_i^m = \delta {\bf u}_i^m \cdot {\tilde {\bf R}}(w_i) \; .
\end{equation}

We now calculate the values of the primitives at the left, $i-1/2$, and right, $i+1/2$, cell interfaces using the primitive variable monotonised differences, 

\begin{eqnarray}
\tilde{{\bf w}}_{L,i+1/2} = {\bf w}_{i} + \left[ \frac{1}{2} - max(\lambda_{i}^{M} , 0) \frac{\delta t}{2\delta x} \right] \delta {\bf w}_{i}^m  \; ,
\label{stoneeqn40}
 \\
\tilde{{\bf w}}_{R,i-1/2} = {\bf w}_{i} - \left[ \frac{1}{2} - max(\lambda_{i}^{0} , 0) \frac{\delta t}{2\delta x} \right] \delta {\bf w}_{i}^m \; ,
\label{stoneeqn41}
\end{eqnarray}
\noindent
where $\lambda_i^M$ and $\lambda_i^0$ are the largest and smallest eigenvalues based upon the cell centred data.

We can now perform the characteristic tracing step, removing from the quantities calculated in equations (\ref{stoneeqn40}) and (\ref{stoneeqn41}), contributions from each set of waves that travelled towards the interface but did not reach it during the forward half-time-step $\delta t/2$ using \citep{Colella1984,Colella1990},

\begin{eqnarray}
{\bf w}_{L,i+1/2} = \tilde{{\bf w}}_{L,i+1/2} + \frac{\delta t}{2\delta x}\sum_{{\lambda}^{\alpha} > 0} \left[ (\lambda_i^M - \lambda_i^{\alpha}){\tilde {\bf L}}^{\alpha} \cdot \delta {\bf w}_i^m \right] {\tilde {\bf R}}_i^{\alpha} \; ,
\label{stoneeqn42}
 \\
{\bf w}_{R,i-1/2} = \tilde{{\bf w}}_{R,i-1/2} + \frac{\delta t}{2\delta x}\sum_{{\lambda}^{\alpha} < 0} \left[ (\lambda_i^0 - \lambda_i^{\alpha}){\tilde {\bf L}}^{\alpha} \cdot \delta {\bf w}_i^m \right] {\tilde {\bf R}}_i^{\alpha} \; .
\label{stoneeqn43}
\end{eqnarray}

With time averaged primitive values calculated for each cell face, Godunov's fluxes can now be computed and applied using a conventional Riemann solver technique such as the accurate Roe scheme. If using an approximate Riemann solver such as the HLL scheme, then due to the averaging of the intermediate eigenstates between the fastest left and right waves, it is necessary to add the additional terms (\ref{stoneeqn44}) and (\ref{stoneeqn45}) shown below, to equations~(\ref{stoneeqn42}) and (\ref{stoneeqn43}) respectively. 

\begin{eqnarray}
{\bf w}_{L,i+1/2} = - \frac{\delta t}{2\delta x}\sum_{{\lambda}^{\alpha} < 0} \left[ (\lambda_i^{\alpha} - \lambda_i^{M}) {\tilde {\bf L}}^{\alpha} \cdot \delta {\bf w}_i^m \right] {\tilde {\bf R}}_i^{\alpha} \; ,
\label{stoneeqn44}
 \\
{\bf w}_{R,i-1/2} = - \frac{\delta t}{2\delta x}\sum_{{\lambda}^{\alpha} > 0} \left[ (\lambda_i^{\alpha} - \lambda_i^{0}){\tilde {\bf L}}^{\alpha} \cdot \delta {\bf w}_i^m \right] {\tilde {\bf R}}_i^{\alpha} \; .
\label{stoneeqn45}
\end{eqnarray}

These remove the effect of waves which propagate away from the cell interface which would otherwise be included, allowing for the scheme to achieve second order accuracy.

Through the use of operator splitting, we can deal with the addition of extra source and sink terms, including gravity, separately after the collisionless stellar hydrodynamics step. To ensure that the gravitational acceleration term is kept second order accurate in time, we store the acceleration from the previous time step and interpolate forward to account for temporal variation in the gravitational field. Additional source and sink terms can include star formation and supernova, although we only consider gravity in this paper.



%

\section{Tests}
\label{sec:tests}

Notwithstanding a formal similarity of the CHE and BME, testing the latter
is nontrivial as it requires knowing analytical solutions for the dynamics of
{\it collisionless} systems. 
Fortunately, some of the test problems that are used to benchmark the performance
of hydro codes can be adapted to test the collisionless systems. 
This is because equations~(\ref{contin})-(\ref{dispers}) can be made formally identical 
to those of classical hydrodynamics in special cases.
Indeed, for a one-dimensional flow of a non-gravitating collisionless fluid 
along the $x$-direction, the BME become
\begin{eqnarray}
\label{contin1D}
\pder{\rho}{t} &+& \pder{}{x}\left( \rho u_{\rm x} \right) =0 \; , \\
\pder{}{t}\left( \rho u_{\rm x} \right) &+& 
\pder{}{x}\left(\rho u_{\rm x} \cdot u_{\rm x} + \rho \sigma_{\rm xx}^2 \right) = 0 \; , \\
\label{dispers1D}
\pder{E_{\rm xx}}{t} &+& \pder{}{x}\left[ \left( E_{\rm xx} + 2 \rho \sigma_{\rm xx}^2 
\right) u_{\rm x}  \right] = 0 \; .
\end{eqnarray}
If we now assume that $\rho \sigma_{xx}^2$ is equal to the gas pressure $P$
in classic hydrodynamics and $\varepsilon=\frac{1}{2}\rho\sigma_{xx}^2$ then as $P=(\gamma-1)\varepsilon=2\varepsilon$, i.e. the ratio of specific heats $\gamma$ is equal to 3,
then equations~(\ref{contin1D})-(\ref{dispers1D}) turn into the following form 
\begin{eqnarray}
\pder{\rho}{t} &+& \pder{}{x}\left( \rho u_{\rm x} \right) =0 \; , \\
\pder{}{t}\left( \rho u_{\rm x} \right) &+& 
\pder{}{x} \left(\rho u_{\rm x} \cdot u_{\rm x}\right)  = - \pder{P}{x} \; , \\
\pder{}{t}\left( \varepsilon+\frac{1}{2}\rho u_{\rm x}^2 \right)
 &+& \pder{}{x} \left[ \left(  \varepsilon+\frac{1}{2}\rho u_{\rm x}^2 + P \right) u_{\rm x}  \right] = 0 \; ,
\end{eqnarray}
which are exactly one-dimensional classic hydrodynamics equations.
The fact that the BME for a one-dimensional flow turns into the CHE for $\gamma=3$ 
is not a coincidence but is related to the fact that a one-dimensional flow of a collisionless
fluid has only {\it one} translational degree of freedom $i=1$, unlike 
a collisional fluid which can still be characterised by $i=3$
due to frequent collisions that equipartition the kinetic energy of particles 
between the three translational degrees of freedom. 
This means that in order to properly describe
the one-dimensional flow of a collisionless fluid using the classical hydrodynamics approach,
one needs to assume that the fluid has only one translational degree of freedom, 
i.e. $\gamma=(i+2)/i=3$.
It is straightforward to show that for a two-dimensional flow of a collisionless fluid,
one needs to consider a specific case with $\rho \sigma_{xx}^2=P$ and $\rho \sigma_{yy}^2=P$
and also $\gamma=2$, i.e. a flow of a collisional fluid with two degrees of freedom.

\begin{figure}
  \resizebox{\hsize}{!}{\includegraphics{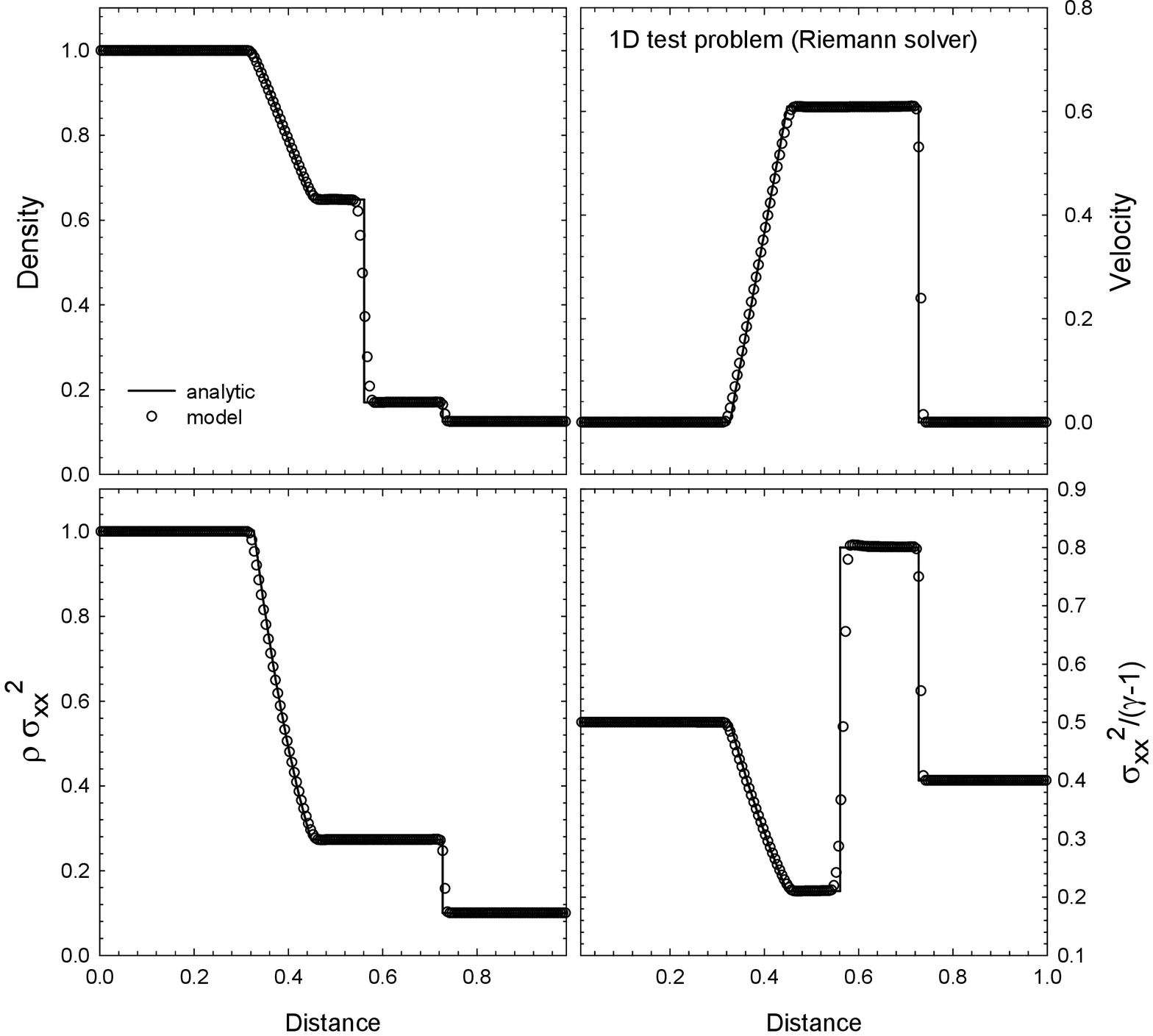}}
      \caption{Comparison of the analytic (solid lines) and numerical (open circles) solutions
      of the one-dimensional Sod shock tube problem. The latter is obtained using the 
      Riemann solver. See the text for more details.}
         \label{fig1}
  \resizebox{\hsize}{!}{\includegraphics{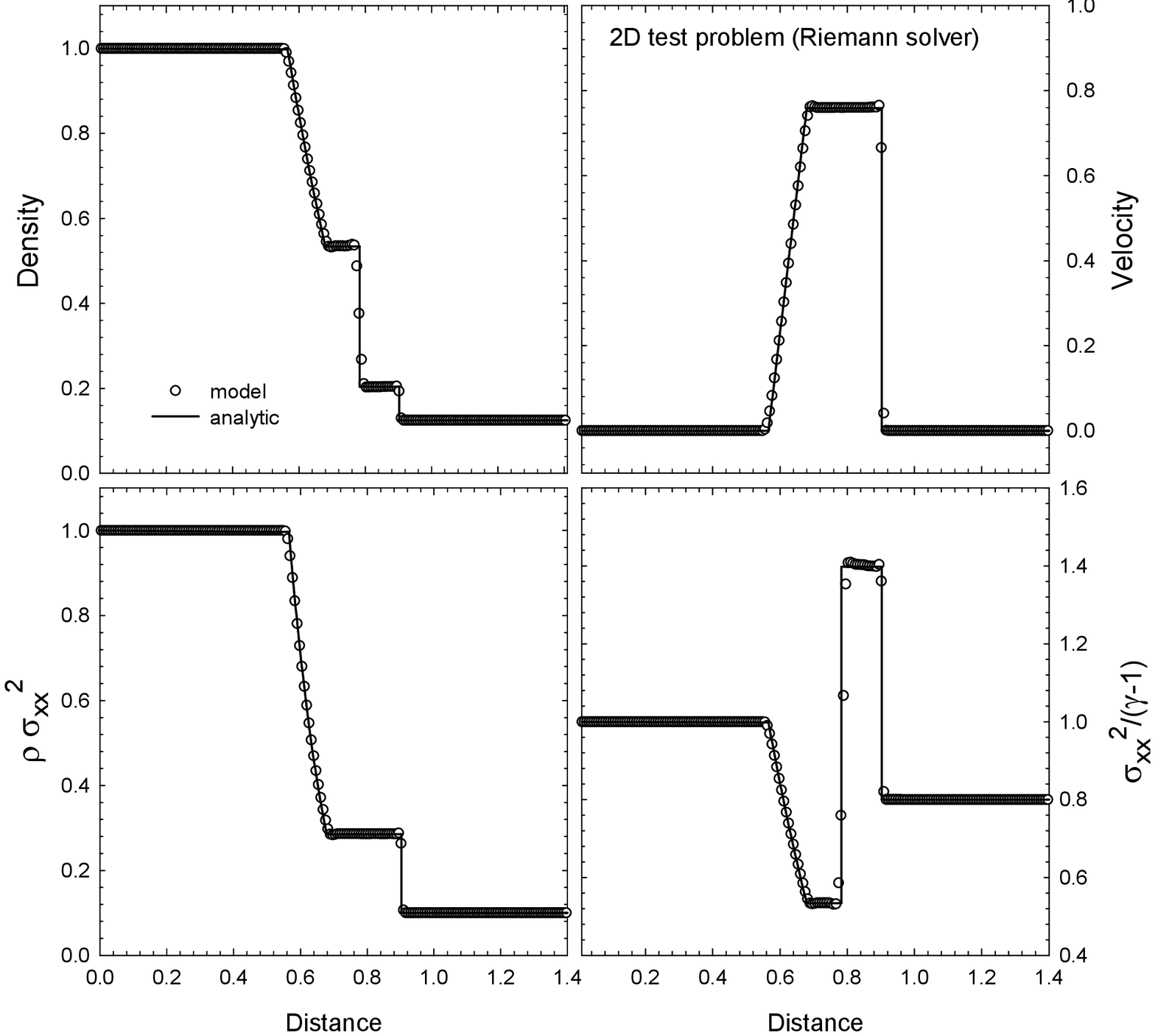}}
      \caption{The same as Fig.~\ref{fig1} only for the two-dimensional Sod shock tube problem.}
         \label{fig2}
         
\end{figure}

\subsection{Sod shock tube}
\label{sec:sodshock}
The above analysis allows us to use the standard Sod shock tube problem, often employed
to benchmark collisional hydro codes. To generate an analytic solution to a one-dimensional
discontinuous state along the $x$-coordinate, we choose $\gamma=3$ and use
the standard solution procedure \citep{Sod78}. For a two-dimensional discontinuous state in
the $x-y$ plane at an arbitrary angle $\alpha$ to the $x$-direction, 
an analytic solution is obtained by choosing $\gamma=2$. 

In the 1D case, we run the test along 
the $x$-coordinate and set $\rho \sigma_{xx}^2$ and $\rho$ 
at $x\in [0-0.5]$ to 1.0, while at $x\in [0.5-1.0]$ the x-component of the stress tensor
is set to 0.1 and the stellar density is 0.125. The numerical resolution is 200 grid zones and the cells are equidistantly spaced.
In the 2D case, we align the initial fluid discontinuity along the [1,1,0] plane and use the same densities and velocity dispersions as were set for the 1D setup setting $\rho\sigma_{xx}^2=\rho\sigma_{yy}^2$ in order to yield agreement with the results for an isotropic collisionally dominated gas. This yields a shock wave which propagates at $45^{\circ}$ to the $x$ and $y$ axes. Again we use 200 cells in the $x$ and $y$ directions and note that the setup can easily be extended to three dimensions although for brevity we do not show these here.

\begin{figure}
  \resizebox{\hsize}{!}{\includegraphics{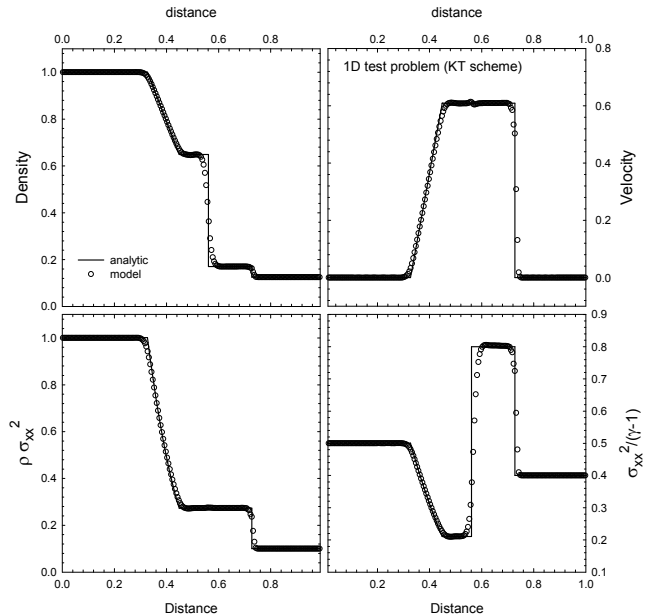}}
      \caption{The same as Fig.~\ref{fig1} only for the KT scheme.}
         \label{fig3}       
\end{figure}

Figures~\ref{fig1} and \ref{fig2} show the results of our tests for the 1D and 2D 
setups, respectively, at $t$=0.1 using the HLLC Riemann solver. 
In particular, the top-left, top-right, bottom-left, and bottom-right
images show the density, velocity, the x-component of the stress tensor $\rho \sigma_{xx}^{2}$, 
and  $\sigma_{xx}^2/(\gamma-1)$. The latter two quantities may be regarded as analogues to
the gas pressure $P$ and internal energy density $\varepsilon$ in classical hydrodynamics.
In the 1D case, the plotted velocity is $u_{\rm x}$, whereas in the 2D case we plot the absolute
value of velocity $\left( u_{\rm x}^2+u_{\rm y}^2 \right)^{1/2}$ along the diagonal $x=y$,
i.e., along the direction of the shock wave propagation. It is evident that the 
Riemann-solver-based numerical scheme can trace the shock waves (propagating to the right) 
very well, usually within one to two grid zones. Contact discontinuities (propagating to the left) are smeared to a greater extent, usually by 5-6 grid zones. 
Nevertheless, the position of the shock waves and contact discontinuities along with the values of the flow variables are reproduced quite accurately.

In a conventional collisional shock, the energy of the macroscopic motion of the gas is transfered through collisions into random isotropic kinetic motions of the particles (thermal energy). In a collisionless fluid, the energy of macroscopic motion is transfered into an effective increase in the velocity dispersion as stellar streams with different velocities mix at the shock front. 

Open circles in Figure~\ref{fig3} present the results of the 1D Sod shock tube test using the KT scheme. It is evident that this scheme does not provide as good an agreement with the analytic solution as the HLLC Riemann solver, particularly for the velocity dispersion (bottom-right panel). In the KT scheme, the shock fronts and contact discontinuities are smeared out over roughly twice as many grid zones as in the HLLC scheme. However encouragingly, the positions of the shocks and discontinuities and also the values of the flow variables are still accurately reproduced. All in all, the HLLC scheme shows an undoubtedly superior performance though at a higher CPU cost, with the KT scheme typically taking around 22\% less time to run and a near identical amount of memory.



\subsection{Collapse of a pressure-free sphere}
\label{sec:sphercollapse}
The gravitational collapse of a pressure-free sphere is used to assess the code's 
ability to accurately treat converging spherical flows on the Cartesian grid.
This test is also useful for estimating the performance 
of the gravitational potential solver on dynamical problems for which one has to
use a finite-difference form of the gravitational potential {\it gradient}.
Since the test setup involves no stress tensor, the resulting equations are 
identical in the collisionless and collisional fluid dynamics cases and 
one can use an analytic solution describing the collapse of every mass shell in
the limit of an infinite sphere radius \citep{Hunter62}.

To run this test, we set a cold homogeneous sphere of unit radius and density (for convenience, the gravitational constant is also set to unity) and let it collapse under its own gravity.  We use a block size of $16^3$ cells and six levels of refinement leading to an effective grid size of $512^3$ cells. We adopt isolated gravitational boundary conditions. Unfortunately, we must consider a cloud of finite radius in Cartesian geometry with a sharp outer boundary to preserve the cloud sphericity. As a result, a rarefaction wave develops after the onset of the collapse, propagating towards the coordinate origin and necessitating complicated corrections to the analytic solution.

\begin{figure}
  \resizebox{\hsize}{!}{\includegraphics{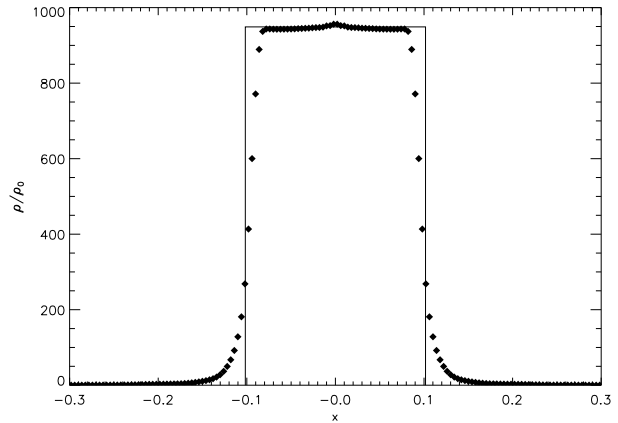}}
      \caption{Comparison of the obtained numerical solution (diamonds) to the analytic solution (solid line) for the collapse of a homogeneous pressureless sphere after 0.985 free fall timescales.}
         \label{fig4}       
\end{figure}

Figure~\ref{fig4} compares the results of our numerical simulation with the ``uncorrected'' analytic solution of \citet{Hunter62} (solid line) at 0.985 free fall timescales, when the initial density has increased by nearly three orders of magnitude. The majority of cells within the sphere lie upon the analytic solution indicating a good homologous collapse with only a small peak above the analytic solution forming within the very central few cells. In addition to this small peak, the initially sharp boundary of the cloud is smeared out over several cells as a result of the rarefaction developing. The analytic radii of the cloud however shows reasonable agreement with the radii at which the half peak density is reached. In general, our results are found to be in strong agreement with those of other authors applying this test problem to the classic hydrodynamics \citep[e.g.,][]{SN92}.

\subsection{Collapse of a sphere with anisotropic stress tensor}
\label{sec:anisocollapse}
As a third test, to demonstrate that the code can model the formation of anisotropies, we initialise two copies of the same sphere as in \S~\ref{sec:sphercollapse} with a non-negligible dispersion tensor. In one we initially set an isotropic dispersion tensor whilst in the second we initially weight the different diagonal terms $\sigma_{xx}^2 : \sigma_{yy}^2 : \sigma_{zz}^2 $ with the ratio of $1:2:3$. We set the system up so that the initial maximum effective dispersion energy is only 25\% of that needed to support the sphere against gravitational collapse. This allows the sphere to collapse down from its initially symmetric configuration and deform into an ellipsoid. The velocity dispersion is then free to change as the simulation evolves. 

\begin{figure}
  \resizebox{\hsize}{!}{\includegraphics{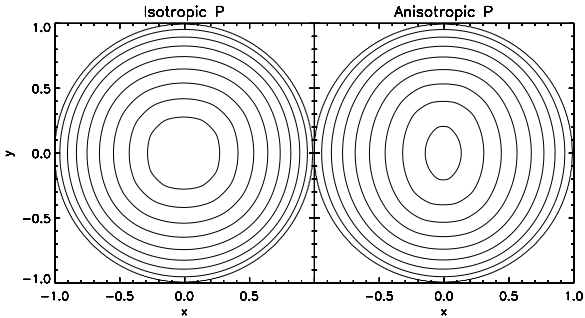}}
      \caption{Plots of the $0.9\rho_{max}$ contour every 0.1 free fall timescales for collapsing spheres of initially homologous, spherical mass distributions. The left hand panel has an isotropic pressure tensor and thus maintains expected sphericity as it collapses, whilst the sphere in the right hand panel deforms over time due to the initial asymmetry in the dispersion tensor, yielding a pronounced ellipse.}
         \label{fig5}       
\end{figure}

We plot in figure~\ref{fig5} the density contour for $\rho = 0.9\rho_{max}$ at regular time intervals for both the cloud with an isotropic dispersion tensor (left hand panel) and an anisotropic dispersion tensor (right hand panel). The isotropic dispersion tensor leads to a collapse which is isotropic as expected whilst in the anisotropic case, the collapse proceeds quickest in the direction with the lowest dispersion term. This leads to a flattening of the sphere into an ellipse.

\begin{figure}
  \resizebox{\hsize}{!}{\includegraphics{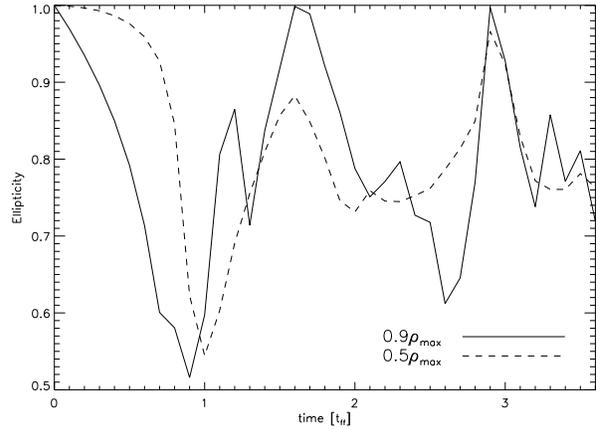}}
      \caption{Ellipticity of the collapsing cloud with an anisotropic dispersion tensor, as a function of time. The ellipticity, the ratio between the minor and major axes, is determined for different iso-density surfaces; $0.9\rho_{max}$ (solid line) and $0.5\rho_{max}$ (dashed line). The major and minor axes begin initially as the z and x axes respectively, but switch after material first reaches maximum compression within the core and rebounds at $t \approx 0.9 t_{ff}$, with material being preferentially expelled along the x-axis.}
         \label{fig6}       
\end{figure}

In figure~\ref{fig6} we plot the ellipticity of the collapsing anisotropic cloud as a function of time. The ellipticity is determined to be the ratio between the minor and major axes of the ellipsoid over a specified iso-density surface. The solid line corresponds to the ellipticity of the iso-density surface of $0.9\rho_{max}$ and the dashed line is the ellipticity for $0.5\rho_{max}$.

Initially, since the velocity dispersion is lowest in the x-direction, material along that axes collapses fastest, with the ellipticity increasing as the density builds. This leads to a greater effective pressure, $P_{xx} \equiv \rho\sigma_{xx}^2$, from the velocity dispersion. Eventually the core collapses to a density and pressure sufficient to stop further collapse. At this point the sphere begins to isotropise as material from the other two axial directions catches up and falls onto the core. This collapse of material onto the core, drives up the pressure and material is forced out around $t \approx 0.9t_{ff}$  along the x-direction where least resistance is offered. This again increases the ellipticity and results in the major and minor axes of the ellipse switching places as the x-axis becomes the major instead of the minor axis and vice versa for the z-axis. The system then oscillates between a high degree of ellipticity and near spherical symmetry as the different axes of the ellipsoid collapse and rebound at different times and speeds. These oscillations repeat over many free fall timescales. Material at a lower density follows the same cyclical pattern of varying ellipticity but with less pronounced maxima and minima. As a result of the non-linear nature of the collapse and its relative separation from the violent oscillations within the core, transitions at lower density are also smoother.

\subsection{Spiral instability in a stellar disc}
\label{sec:diskgal}
The final test problem describes the growth of a spiral structure
in a gravitationally unstable stellar disc.
The stability analysis of a thin stellar disc to a local axisymmetric
perturbation states that the disc is gravitationally unstable if
\citep{Toomre64}
\begin{equation}
Q_{\rm T}= \frac{\sigma_{\rm rr} \kappa}{3.36 G \Sigma} <Q_{\rm c}=1.0 \; ,
\end{equation}
where $\sigma_{\rm rr}$ is the radial stellar velocity dispersion,
$\kappa$ is
the epicycle frequency, and $\Sigma$ is the stellar surface density.
A finite disc thickness and non-axisymmetric perturbations may increase
the critical Toomre parameter $Q_{\rm c}$ by a factor of unity.

The initial setup consists of a self-gravitating, rotating stellar
disc submerged in a fixed dark matter (DM) halo. The latter is
described by the quasi-isothermal density profile of the form
\begin{equation}
\label{qithermal}
\rho_{\rm DM}= \frac{\rho_{\rm 0} }{ 1 + (\varpi/r_{\rm 0})^2} \; .
\end{equation}
where $\varpi$ is the galactocentric distance.
The central density $\rho_{\rm 0}=1.24\times 10^{-2}~M_\odot$~pc$^{-3}$ and
the characteristic scale length
of the quasi-isothermal halo $r_{\rm 0}=3.2$~kpc are calculated based on the
assumed DM halo mass $M_{\rm DM}=2\times 10^{11}~M_\odot$ and halo radius,
$r_{\rm h} = 124.7$~kpc, using relations provided in \citet{VRH2012}.

\begin{figure}
  \resizebox{\hsize}{!}{\includegraphics{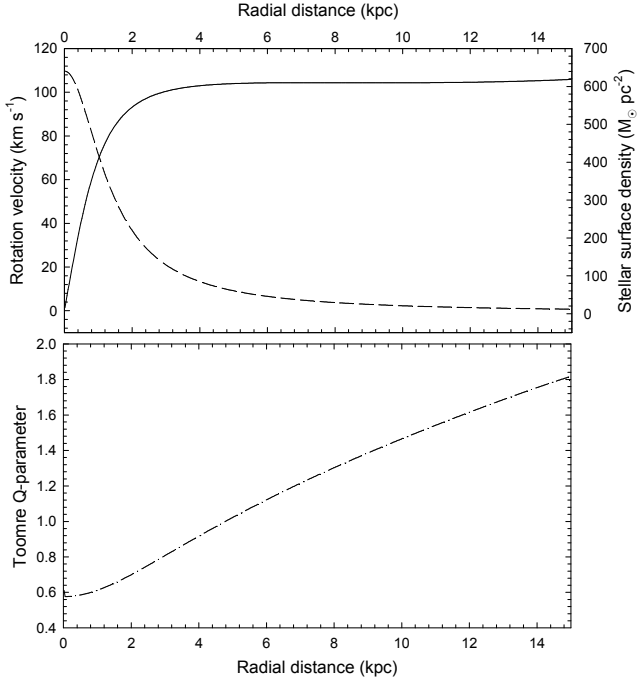}}
      \caption{Initial radial distributions of the rotation velocity
(solid line), stellar surface
      density (dashed line) and Toomre $Q$-parameter (dash-dotted line) of
the model disc. }
         \label{figSp1}
\end{figure}

The computational box spans 30~kpc in each coordinate direction $(x,y,z)$.
The numerical procedure for generating
a rotationally supported stellar disc in the combined gravitational
potential of stars and DM halo is described in detail in \citet{VRH2012}.
The rotation curve, the radial gas surface density profile and the
radial profile of the Toomre parameter $Q_{\rm T}$ of the initial stellar
disc are plotted in Fig.~\ref{figSp1} with the solid,
dashed, and dot-dashed lines, respectively.
Evidently, the disc is initially gravitationally unstable.

\begin{figure}
\resizebox{\hsize}{!}{
\includegraphics[trim=0.7cm 0.2 1.25cm 0cm, clip=true]{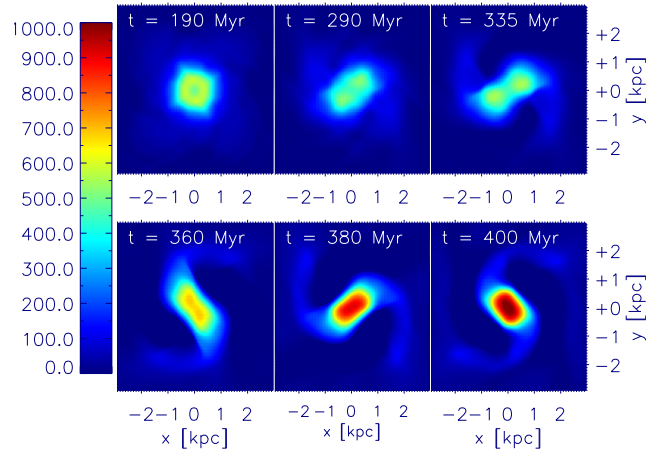}}
      \caption{Disc images showing the growth a a two-armed spiral pattern 
        over time. For each snapshot, the difference between the initial 
        azimuthally symmetric disk surface density profile and the current 
        profile is shown. The colour bar is in units of $M_{\odot} pc^{-2}$
        and the time is shown in the top right of each figure, 
        corresponding to 190, 290, 335, 360, 380 and 400 Myrs. }
         \label{figSp2}
\end{figure}

Figure~\ref{figSp2} presents a series of disc images at several successive
times
since the beginning of numerical simulations. Usually, an initial seed
perturbation
is required to drive the system out of equilibrium and initiate the growth of
a spiral structure. However, in our case, the initial perturbation is
introduced
due to a re-map of the initial equilibrium configuration from the
cylindrical coordinates
\citep{VRH2012} onto the Cartesian coordinates (the FLASH code). The
growth of a two-armed
spiral mode is apparent in the figure.

To quantify the growth rate of spiral modes in our model,
we employ global Fourier amplitudes defined as
\begin{equation}
A_{\rm m}(t) = \frac{1}{M_{\rm d}} \left| \int \limits_{-X} \limits^{+X}
\int \limits_{-Y}
\limits^{+Y} \Sigma(x,y,t) e^{i m \phi} dx dy  \right| \; ,
\end{equation}
where $m$ is the spiral mode, $\Sigma(x,y,t)$ is the surface density of
the stellar disc,
$\phi=\arctan({y/x})$ is the polar angle, and $M_{\rm d}$ is the total
mass of the
stellar disc. The amplitudes are calculated on a square region with size
$(-X:X,-Y:Y)=(-2:2,-2:2)$~kpc,
which is centred on the coordinate origin and encompasses the growing
spiral.
The global Fourier amplitudes can be regarded as the amplitude of a spiral density perturbation
relative to the surface density of the axisymmetric disc.

\begin{figure}
  \resizebox{\hsize}{!}{\includegraphics{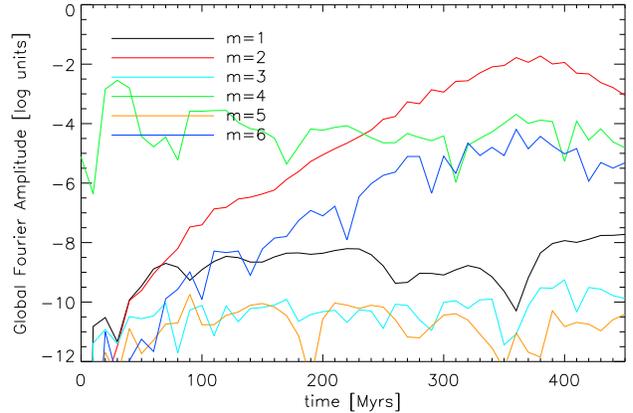}}
      \caption{Global Fourier amplitudes as function of time elapsed since
the beginning of numerical simulation. }
         \label{figSp3}
\end{figure}

Figure~\ref{figSp3} shows the temporal evolution of the
first six global Fourier amplitudes (in log units) in our model
stellar disc. Initially all amplitudes are negligibly small except for that
corresponding to the $m=4$ mode. This is a numerical artifact and arises due to the 
finite resolution within the core. Similar to the results of \citet{Tasker2008} in which they 
model a static King-profile that is initially in hydrostatic equilibrium, the 
finite resolution within the core leads to an under-resolved gravitational potential. This 
results in a slight relaxation and expansion of the very central region. An outward flow of
material results through the faces of the central most cell and due to the fact 
that our disk is constructed so as to promote the growth of gravitational instabilities, 
this generates a small yet non-negligible perturbation. Although similar behaviour may occur 
in all mesh codes, those using Cartesian geometries are particularly susceptible given their 
strong natural bias for flows along the axes of the mesh. Despite being present from the very 
onset of the simulation, the $m=4$ mode slowly decays and does not lead to the growth of a 
spurious four armed spiral given the dominance of the $m=2$ mode at late times by around
two orders of magnitude.  

The $m=2$ mode demonstrates the fastest growth rate. It saturates at
around $t\approx370$~Myr and
declines slowly in the subsequent evolution due to continuing disc heating.
The maximum amplitude of the $m=2$ mode is about $10^{-1.73}$, which suggests
a non-axisymmetric density perturbation of about 2\% relative
to the underlying axisymmetric disc.

An appealing physical interpretation
for the growth of a spiral structure in our model disc
is swing amplification. Amplification occurs when any leading spiral
disturbance, e.g., introduced by the initial seed perturbation,
unwinds into a trailing one due to differential rotation
\citep[e.g.][]{Toomre81,Athanassoula84,VT2006,VT2008}.
In addition, swing amplification needs a feedback
mechanism that can constantly feed a disc with leading spiral disturbances.
Trailing short-wavelength disturbances propagating
through the disc centre and emerging on the other side as leading ones
provide a feedback for the swing amplifier \citep{Toomre81}.

\begin{figure}
  \resizebox{\hsize}{!}{\includegraphics{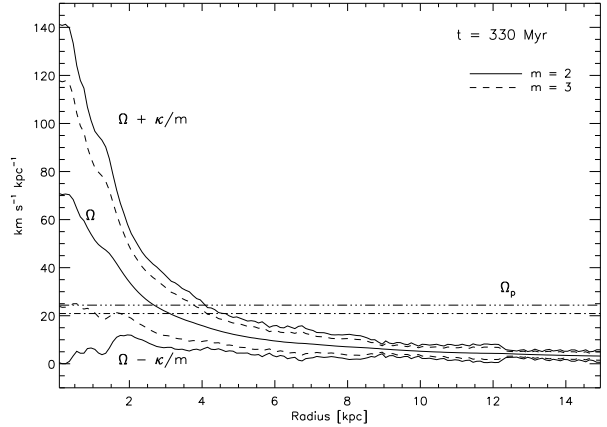}}
      \caption{Radial behaviour of $\Omega$, $\Omega \pm \kappa/m$, and $\Omega_{\rm p}$ in our model stellar disc, where $\Omega$ is the azimuthally averaged angular velocity of stars and $\Omega_{\rm p}$ is the angular speed of the spiral pattern. The dash-dot-dotted and dash-dotted lines show $\Omega_{\rm p}$ as measured at two distinct positions in the disc; 2.0~kpc and 4.0~kpc respectively. The intersections of $\Omega \pm \kappa/m$ and $\Omega_{\rm p}$ determine the positions of Lindblad resonances for the corresponding spiral mode $m$. 
}
       \label{figSp4}
\end{figure}

We can check if the growth mechanism of our spiral structure is consistent
with the predictions of the swing amplification theory.
For our spiral pattern to be triggered by the swing amplifier,
there should be no inner Lindblad resonance (ILR) for the $m=2$ mode.
Otherwise, the corresponding trailing disturbances will damp at the ILR,
thus failing to pass through the disc centre and promoting the growth
of the $m=2$ mode. Figure~\ref{figSp4}
shows the radial profiles of $\Omega$ and $\Omega\pm \kappa/m$ at
$t=330$~Myr, where
$\Omega$ is the angular velocity of the stellar disc and $\kappa$
is the epicycle frequency. Both quantities are the azimuth averages.
In particular, the solid and dashed lines show the corresponding
values for the $m=2$ and $m=3$ modes, respectively. The dash-dot-dot-dotted and
dash-dotted lines show the angular velocity $\Omega_{\rm p}$ of the global
spiral pattern measured at two distinct positions in the disc: 2.0~kpc and
4.0~kpc.
In theory, $\Omega_{\rm p}$ should be independent of radius,
though in practice it is always slightly faster in the inner regions than
in the outer ones due to gradual winding of a spiral pattern.
The radial position of Lindblad resonances  are determined as
$m(\Omega-\Omega_{\rm p})= \pm \kappa$,
i.e., as the radial distance where the forcing frequency of the spiral
pattern
coincides with the epicycle frequency of the stars.
In particular, the inner and outer Lindblad resonances correspond
to the plus and minus signs, respectively.

Figure~\ref{figSp4} demonstrates that there is clearly no ILR for the $m=2$ mode,
whereas there is one marginally for the $m=3$ mode (and thus for any higher-order mode).
Indeed, the dash-dotted and dash-dot-dot-dotted lines never intersect the $\Omega-\kappa/2$ curve, whereas
they do intersect the $\Omega - \kappa/3$ curve at $r<1$~kpc, thus damping
$m=3$ disturbances that try to pass through the disc center. Although slight growth over time 
of the $m=3$ mode is observed in figure~\ref{figSp3}, it is damped relative to the $m=2$ mode
by over two orders of magnitude. Higher order modes for which a clearer
ILR will be present are damped to a much greater extent.
This explains why the $m=2$ mode (and not any higher-order mode) ultimately dominates the spiral pattern.

Another consistency check on our model is the position of the outer
Lindblad resonance (OLR).
Spiral disturbances cannot propagate through the OLR, which effectively
limits the radial
extent of a spiral pattern.
Figure~\ref{figSp4} indicates that the OLR for the $m=2$ mode is
located approximately at 4~kpc and this value is in agreement with the radial
extent of the spiral pattern in Fig.~\ref{figSp2}.
We conclude that our numerical results are in general agreement with
the predictions of the swing amplification theory.

\section{Discussion and Summary}
\label{sec:summary}
In this paper we have outlined a series of important limitations that are encountered in mesh codes when using the particle-mesh technique. Used to consistently integrate collisionless stellar or dark matter components into simulations which model gas physics on an Eulerian mesh, the particle-mesh technique has the advantage of being relatively straight forward to implement but is unfortunately subject to several key limitations.  From a technical perspective these include poor load balancing and increased communication overhead. From a physical point of view, discreteness effects incurred when discrete particle properties are mapped to the mesh, are believed to result in spurious entropy generation. In galaxy formation simulations, this can have a significant influence on the reservoir of cold gas available to form stars, potentially changing the entire dynamics of the system.

To overcome these limitations we have proposed the use of the collisionless Boltzmann moment equations as a powerful alternative. Such an approach allows us to model the collective properties of collisionless objects such as stars as a fluid, instead of relying upon the traditional N-body approach. In this paper we refer to this approach as ``collisionless stellar hydrodynamics'' given its primary use in the modelling of the properties of large collections of stars, such as those in a galaxy, in much the same way as we model gases in Eulerian grid codes.
Although direct integration of the collisionless Boltzmann equation has been undertaken by other authors, it is often prohibitively expensive to perform. This arises as it is necessary to track both the spatial domain and its corresponding velocity phase space. However \citet{Yoshikawa2012} demonstrates that the ability to resolve small scale velocity fluctuations is much greater in these codes than for conventional N-body simulations.

Instead of a costly direct integration of the Boltzmann equation, we opt to derive the zeroth, first and second order moments under the usual zero heat flux approximation and taking the third order moments to be negligible. Although ultimately an approximation, we show that this yields results in-line with a series of demanding test problems, requiring a factor of $32^3$ less memory than is needed using direct integration over the complete velocity phase space.

We implement these new collisionless stellar hydrodynamic equations into the massively parallel \flash\ \amr\ code using two different numerical solvers for hyperbolic partial differential equations. The first scheme by \citet{KurganovTadmor2000}, requires no information on the equation of state of a collisionless fluid or any assumptions about the behaviour of fluid interactions within the Riemann fan. This Riemann solver free prescription allows us to track sharp discontinuities with relatively low numerical diffusion and provides a very useful comparison with which to compare our more sophisticated second scheme. In future, its extensibility is hoped to be used to explore the impact of including the higher order moments which we exclude here.

By assuming a diagonalised dispersion tensor, we have derived the eigenvalues and eigenvectors for the first three moments of the collisionless Boltzmann equations. These have been implemented into a characteristic tracing method based on the MUSCL-Hancock unsplit advection scheme. This allows time averaged primitive states to be calculated at the cell interfaces using characteristic tracing, from which time averaged fluxes can be computed using a range of Riemann solvers.  

We validate our code using a suite of tests with analytic solutions, including some commonly used to benchmark conventional hydrodynamic codes. By realising that the classic hydrodynamic equations and Boltzmann moment equations can be reconciled under certain conditions, we are able to generate analytic solutions for the Sod shock test which are applicable for collisionless systems. Of particular note, we find that since a collisionless fluid has only one degree of freedom in a given direction and is unable to equipartition the kinetic energy of particles between the three dimensions through collisions, it has an effective ratio of specific heats, $\gamma = 3$. Thus in 1D, a Sod shock solution for a collisionally dominated gas with $\gamma = 3$, matches that for a collisionless fluid. Using this insight, we confirm that our numerical schemes; both the \ktscheme\ and characteristic tracing method, accurately reproduce analytic solutions. Although the \ktscheme\ runs 22\% faster with near identical memory consumption, our Riemann solver based scheme shows notably better resolving power of sharp shocks and discontinuities, spreading them over roughly half the number of cells.

Through the use of a spherical pressure-free collapse problem, we confirm that the code can maintain good flow symmetry in convergent flows in the absence of any anisotropy in the dispersion tensor. Agreement with the peak density within the collapsed sphere also confirms the validity of the Poisson solver used in \flash. Through the inclusion of a non-negligible anisotropic dispersion tensor, we extend the spherical collapse to observe the behaviour of an initially homologous spherical cloud as it collapses and deforms into an elliptical profile. Although initial behaviour of the system occurs in-line with expectations, the long term behaviour is more complicated with repeated transitions between an elliptical and spherical profile, along with switching of the major and minor axes after the initial core implosion. Although the inclusion of gas physics may help to isotropise the internal structure of the cloud and damp these oscillations, they will nevertheless remain important to the dynamics outside of the core. We will explore in more detail interactions between the collisionally dominated gas physics and the stellar material in future papers. In particular, we will extend our scheme to include off-diagonal terms and a more thorough exploration of the effects of higher order terms. Although powerful in its current form, the addition of off-diagonal terms in the velocity dispersion tensor will allow us to directly measure the vertex deviation, a key observable used for probing galactic structure. To date observers have made extensive measurements of our own Milky Way's vertex deviations which can hide detailed information about the kinematic properties of the bar and spiral arms. However noise in particle based simulations has made accurate theoretical predictions difficult, something which our new collisionless stellar hydrodynamics code can allow us to overcome. Having already confirmed the applicability of the collisionless stellar hydrodynamics code to the formation of spiral structure, we have found excellent agreement between our numerical simulations of the growing spiral pattern and the predictions of the swing amplification theory, with the correct relation between the disc size and the outer Lindblad resonances being observed.

As a final note, we find the time taken in \flash\ for the Riemann solver based collisionless stellar hydrodynamics routine is only 7\% greater than that for the original classic hydrodynamic scheme, allowing the collisionless stellar hydrodynamics to scale linearly with the classic hydrodynamics. We also find that instead of doubling the overall amount of time spent communicating when both gas and collisionless stellar hydrodynamics are included, our grouping of communications results in a net increase of only 1.2\% to the overall communication time compared to that of just a single phase hydrodynamic scheme. Thus we conclude by confirming that our new collisionless stellar hydrodynamic approach using the Boltzmann moment equations both preserves the excellent scaling previously demonstrated in \flash\ as well as enhancing the level of detail we can expect to be able to extract from our results, with none of the discreteness effects observed for particle-mesh techniques.

\section*{Acknowledgements} 

The authors acknowledge the generous support of the Austrian FWF for project numbers P21097-N16 (Nigel L. Mitchell) and Lise Meitner fellowship M1255-N16 (Eduard I. Vorobyov). The software used in this work was in part developed by the DOE NNSA-ASC OASCR Flash Center at the University of Chicago. The computations were performed at the Vienna Scientific Cluster (VSC) and the AstroCluster of the Institute for Astrophysics, funded by the University of Vienna.

\appendix

\section{Convolving the Boltzmann moment equations to the classic hydrodynamics form}
On general grounds one can argue that the Boltzmann moment equations (\ref{contin})-(\ref{dispers})
should convolve to the classic hydrodynamics equations (CHE) 
if the stress tensor $\Pi_{ij}$ is diagonal and isotropic. In this section we demonstrate
this by assuming that
$\Pi_{ij}=0$ for $i\ne j$ and $\Pi_{\rm xx}=\Pi_{\rm yy}=\Pi_{\rm zz}$, 
implying that $\sigma_{\rm xx}=\sigma_{\rm yy}=\sigma_{\rm zz}=\sigma$. 
The equation of continuity (\ref{contin}) is identical to that of the CHE and 
the momenta equations (\ref{moment}) trivially turn into their collisional counterparts
with a substitution $\Pi_{\rm xx}=\rho \sigma_{\rm xx}^2=\rho \sigma^2=P$ (and identical
relations for other components), where $P$ is the gas pressure. 

It takes a bit more math to show that the dispersion equations~(\ref{dispers}) can be convolved
to the equation for the total energy per unit volume $E_{\rm T}=\varepsilon +\rho |u|^2/2$.
Let us expand equation~(\ref{dispers}) into the component form in Cartesian coordinates. 
\begin{eqnarray}
\label{eqn1}
\pder{E_{\rm xx}}{t} &+& \pder{}{x_k}\left( E_{\rm xx} u_k +2 \rho \sigma_{{\rm x}k}^2 u_{\rm x} \right)
- 2 \rho u_{\rm x} a_{\rm x} =0 \; , \\
\pder{E_{\rm yy}}{t} &+& \pder{}{x_k}\left( E_{\rm yy} u_k + 2 \rho \sigma_{{\rm y}k}^2 u_{\rm y }\right)-
2 \rho u_{\rm y} a_{\rm y} =0 \; , \\
\label{eqn3}
\pder{E_{\rm zz}}{t} &+& \pder{}{x_k}\left( E_{\rm zz} u_k +2 \rho \sigma_{{\rm z}k}^2 u_{\rm z }\right)-
2 \rho u_{\rm z} a_{\rm z} = 0 \; . 
\end{eqnarray} 
After summing up equations~(\ref{eqn1})-(\ref{eqn3}) we obtain
\begin{eqnarray}
\pder{}{t} \Big( \frac{3}{2} \rho \sigma^2 &+& \frac{1}{2}\rho |u|^2 \Big) + \nonumber \\ 
&+& \pder{}{x_k} \left[ u_k \left( \frac{3}{2} \rho \sigma^2 + \frac{1}{2} \rho |u|^2 \right) 
+ \rho \sigma_{ik}^2 u_i  \right] \nonumber \\ 
\label{eqn4}
&-& \rho u_i a_i = 0 \; ,
\end{eqnarray}
where we have divided the resulting equation by a factor of 2 and have taken into 
account that $\sigma_{\rm xx}=\sigma_{\rm yy}=\sigma_{\rm zz}=\sigma$
and $|u|^2=u_{\rm x}^2+u_{\rm y}^2+u_{\rm z}^2$. In the classical hydrodynamics
the gas pressure relates to the internal energy density via the relation
$P=\varepsilon (\gamma -1)$, where $\gamma=5/3$ for a flow with three
translational degrees of freedom, meaning that $\frac{3}{2} \rho \sigma^2=\varepsilon$ in the
above equation. The final step is to note extract the diagonal part from the symmetric tensor 
$\rho \sigma_{ik}^2$ by introducing the viscous stress tensor $\pi{ik}$.
\begin{equation}
\rho \sigma_{ik}^2 = P\delta_{ik} - \left(P \delta_{ik} - \rho \sigma_{ik}^2 \right) =
P\delta_{ik} - \pi_{ik} \; .
\end{equation}
With this final transformation, equation~(\ref{eqn4}) turns into the usual hydrodynamics
equation for the total energy per unit volume $E_{\rm T}$ of a viscous fluid.
\begin{equation}
\label{eqn5}
\pder{E_{\rm T}}{t} + \pder{}{x_k} \left[ u_k \left(E_{\rm T} + P \right) - \pi_{ik} u_i  \right] -
\rho u_i a_i =0 \; .
\end{equation}

\section{Zero-heat-flux approximation}
When deriving the Boltzmann moment equations, each equation for the $n$-th moment
inevitably requires the knowledge of the $(n+1)$ moment. This chain is 
conventionally closed by setting the third-order moments to zero,
i.e., $Q_{ijk}=\rho^{-1}\int f\,(v_i-u_i)(v_j-u_j)(v_k-u_k)\,d^3{\bl v}=0$.
In this section, we discuss the validity of this approximation.

For this purpose it is useful to refer to a similar problem in the collisional
hydrodynamics, for which the closure problem exists as well. Indeed, the CHE are
derived from the {\it collisional} Boltzmann equation and
if the zero-heat-flux approximation were relaxed, then equation~(\ref{eqn5})
would have an additional term 
$$\frac{1}{2}\pder{}{x_k}\left(\rho Q_{ijk}\right) \; .$$
A contemporary  swindle is to use the empirical Fourier law for thermal conduction and
assume that 
$$ \frac{1}{2}\rho Q_{ijk}=-k(T)\pder{T}{x_k} \; ,  $$
where $T$ is the gas temperature and $k(T)$ is the heat conduction coefficient.
Unfortunately, such a trick cannot be applied to a collisionless fluid as it, strictly
speaking, has no temperature due to the lack of local thermal equilibrium.

However, this analysis can give us an insight as to when the zero-heat-flux
approximation cannot be neglected. In the CHE, such a situation arises 
on a quasi-stationary contact discontinuity between two 
gases with vastly different temperatures when
the thermal diffusion time scale $\tau_{\rm th}=\rho L^2/k(T)$ can be much shorter than 
dynamical one $\tau_{\rm dyn}=L/|u|$, where $L$ is the characteristic size of the system. 
In most astrophysical environments, however, the bulk motions
of gas ensure that $\tau_{\rm dyn}\ll\tau_{\rm th} $ and the third-order moments can be neglected.

In the case of collisionless systems, the diffusion time scale is played by the
typical relaxation time $\tau_{\rm rel}=L/\sigma$, which should be compared against $\tau_{\rm dyn}$.
It is apparent that the zero-heat-flux approximation is expected to be valid if
\begin{equation}
\frac{\tau_{\rm rel}}{\tau_{\rm dyn}} = \frac{|u|}{\sigma } \gg 1 \; .
\end{equation}
This condition is usually met in any rotationally supported system like a galactic stellar disc, with 
\citet{Hensler1995} highlighting stellar dynamic models which show the third 
order moments decay faster than the relaxation timescales. This means that the zero-heat-flux
approximation is approached faster than isotropisation.
However, this condition may break in globular clusters that are thought to be supported against 
gravitational collapse by random motions of stars (i.e., by high velocity dispersions).
This does not invalidate the whole Boltzmann moment approach approach, but 
simply means that an important piece of physics is missing and needs to be taken into account.
We work on developing useful approximations that can treat such cases in collisionless
fluids.


\end{document}